\documentstyle[aps,pre,floats,epsfig]{revtex}

\newcommand {\gs}{\
  \raisebox{-0.2ex}{$\stackrel{\scriptstyle>}{\scriptstyle\sim}$}\ }
\newcommand{\ls}{\
  \raisebox{-0.2ex}{$\stackrel{\scriptstyle<}{\scriptstyle\sim}$}\ }
\begin{document}
\twocolumn[\hsize\textwidth\columnwidth\hsize\csname@twocolumnfalse%
\endcsname
\title{3D Spinodal Decomposition in the Inertial Regime}
\author{V. M. Kendon$^1$, J-C. Desplat$^2$, P. Bladon$^1$ and M. E.
Cates$^1$}
\address{$^1$Department of Physics and Astronomy, 
University of Edinburgh, JCMB King's Buildings, Mayfield Road, Edinburgh\\
EH9 3JZ, United Kingdom\\
$^2$Edinburgh Parallel Computing Centre,
University of Edinburgh, JCMB King's Buildings, Mayfield Road, Edinburgh\\
EH9 3JZ, United Kingdom}
\maketitle
\vspace{-0.4cm}
\center{\small{(Received 25 February 1999)}}
\begin{abstract}
We simulate late-stage coarsening of a 3D symmetric binary
fluid using a lattice Boltzmann method.
With reduced lengths and times, $l$ and $t$ respectively
(with scales set by viscosity, density and surface tension)
our data sets cover $1\ls l \ls 10^5$, $10 \ls t \ls 10^8$.
We achieve Reynolds numbers approaching $350$.
At Re $\gs 100$ we find clear evidence of Furukawa's inertial scaling ($l\sim
t^{2/3}$),
although the crossover from the viscous regime ($l\sim t$) is very broad.
Though it cannot be ruled out, we find no indication that Re is self-limiting
($l\sim t^{1/2}$) as proposed by M. Grant and K. R. Elder\linebreak
{$\left[\right.$}Phys. Rev. Lett.  {\bf 82}, 14 (1999){$\left.\right]$}.\hfill
{\rule[0ex]{4ex}{0ex}}
\end{abstract}
\bigskip
\footnotesize{\hspace{2cm}PACS numbers: 64.75+g, 07.05.Tp, 82.20.Wt}\hfill
\bigskip
]
When an incompressible binary fluid mixture is quenched far below
its spinodal temperature, it will phase separate into domains of
different composition. Here we consider only fully symmetric 50/50 mixtures
in three
dimensions, for which these domains will, at late times, form a bicontinuous
structure, with
sharp, well-developed interfaces. The late-time evolution of this
structure remains incompletely understood despite
theoretical \cite{siggia,furukawa,bray,grant}, experimental
\cite{experiments} and simulation
\cite{laradji,lebowitz,jury,footlin,foot2_3}
work.

As emphasized by Siggia \cite{siggia} and Furukawa \cite{furukawa},
the physics of spinodal
decomposition involves capillary forces, viscous dissipation, and
fluid inertia. Thus, assuming that no other physics enters,
the control parameters are interfacial tension
$\sigma$, fluid mass density $\rho$, and shear viscosity $\eta$. From these
can be
constructed only one length, $L_0 =
\eta^2/\rho\sigma$ and one time
$T_0 = \eta^3/\rho\sigma^2$. We define the lengthscale $L(T)$ of
the domain structure at time $T$ via the structure factor $S(k)$ as
$L = 2\pi  \int S(k) dk /\int k S(k) dk$.
The exclusion of other physics in late stage growth then leads us to the
dynamical scaling hypothesis
\cite{siggia,furukawa}:
$l = l(t)$,
where we use reduced time and length variables, $l \equiv L/L_0$
and $t \equiv (T-T_{int})/T_0$.  Since dynamical scaling should hold only after
interfaces have become sharp, and transport by molecular diffusion
ignorable, we have allowed for a nonuniversal offset $T_{int}$;
thereafter the scaling function $l(t)$ should
approach a universal form, the same for all (fully symmetric,
deep-quenched, incompressible) binary fluid mixtures.

It was argued further by Furukawa \cite{furukawa} that, for small enough
$t$, fluid inertia is negligible compared to viscosity, whereas for large
enough $t$ the reverse is true. Dimensional
analysis then requires the following asymptotes:
\begin{equation}
l \to b t \mbox{;\hspace{3em}} t\ll
t^* \label{viscous}
\end{equation}
\begin{equation}
l \to c t^{2/3} \mbox{;\hspace{3em}} t\gg t^* \label{inertial}
\end{equation}
where, if dynamical scaling holds, amplitudes $b,c$
(and the crossover time $t^*$) are universal.
The Reynolds number, conventionally defined as,
${\mathrm Re} = \rho/\eta L dL/dT = l\dot l$,
%\begin{equation}
%{\mathrm Re} = \frac{\rho}{\eta} L \frac{{\mathrm d}L}{{\mathrm d}T} = l\dot l,
%\label{reynum}
%\end{equation}
becomes indefinitely large in the inertial regime, Eq. (\ref{inertial}).

In a recent paper, Grant and Elder have argued
\cite{grant} that the Reynolds number cannot, in fact, continue to grow
indefinitely. If so, Eq. (\ref{inertial}) is not truly the large $t$
asymptote,
which must instead have $l\sim t^\alpha$ with $\alpha \le \frac{1}{2}$.
Grant and Elder argue that at large enough Re, turbulent remixing of the
interface
will limit the coarsening rate \cite{grant}, so that Re stays bounded. A
saturating Re
(which they estimate as Re $\sim 10-100$) would require any  $t^{2/3}$
regime to eventually cross over to a limiting $t^{1/2}$ law. But if a
single length
scale $l\sim t^{1/2}$ is involved, a saturating Re implies
balance between viscous and inertial terms ($t^{-3/2}$),
while the driving term (interfacial
tension) remains much larger than either ($t^{-1}$). This suggests a failure of
scaling altogether, with at least two length scales relevant at late times.
In any
case, the arguments of Grant and Elder are far from rigorous;
the coarsening interfaces could, remain one step ahead of the
remixing
despite an ever-increasing Re which, if applied to a {\em static}
interfacial structure, would break it up. Thus
Eq. (\ref{inertial})
cannot yet be ruled out as a limiting law.

In what follows we present the first large-scale simulations of 3D spinodal
decomposition to unambiguously attain a regime in which inertial forces
dominate
over viscous ones. We find direct evidence for Furukawa's
$l\sim t^{2/3}$ scaling, Eq. (\ref{inertial}). Although a further crossover to
a regime
of saturating Re cannot be ruled out, we find no evidence for this up to Re
$\simeq
350$.  Our work, which is of unprecedented scope,
also probes the viscous scaling regime
$[$Eq. (\ref{viscous})$]$,
and the nature of the crossover between this and Eq. (\ref{inertial}).
Full details of our results
\cite{long} and of the simulation algorithm \cite{ludwig} will be published
elsewhere.

Our simulations use a lattice Boltzmann (LB) method \cite{higuera,swift}
with the following model free energy:
\begin{equation}
F = \int d{\mathbf r}\left\{-\frac{A}{2}\phi^2+\frac{B}{4}\phi^4
  + \tilde\rho\ln\tilde\rho +
\frac{\kappa}{2}|\nabla\phi|^2\right\},
\end{equation}
in which $A$, $B$ and $\kappa$ are parameters that determine
quench-depth (${A/B} \to 1$ for a deep quench)
and interfacial tension ($\sigma = \sqrt{8\kappa A^3/9B^2}$);
$\phi$ is the usual order parameter (the normalized difference in number
density of the two fluid species);  $\tilde\rho$
is the
total fluid density, which remains (virtually) constant
throughout \cite{swift,ladd}.

The simulation code follows closely that of \cite{swift} (for details see
\cite{long,ludwig}) and uses a cubic lattice with nearest and next-nearest
neighbor
interactions (D3Q15). It was run on Cray T3D and Hitachi SR-2201 
parallel machines with
system sizes up to $256^3$.  The LB method allows the user to
choose
$\eta,\sigma,\rho$ (we set $\rho = 1$ without
loss of
generality), along with the order-parameter mobility
$M$ defined via
$\dot \phi = \nabla \cdot M \nabla (\delta F/\delta \phi)$.
Although it plays no role in the arguments leading to
Eqs. (\ref{viscous}) and (\ref{inertial}), $M$ must be chosen with some care to
ensure that
at late times (a) the algorithm remains stable, (b) the local interfacial
profiles remain close to equilibrium (so that $\sigma$ is well-defined),
and (c) the direct contribution of diffusion to coarsening is negligibly small.
Table \ref{table:pars} shows the
parameters used for our eight $256^3$ runs.
%(Various smaller runs were also done.)

%\renewcommand{\arraystretch}{1.1}
\begin{table}[htb!]
  \caption{Parameters used in LB runs}
  \begin{center}
  \begin{tabular}{r@{\/}lr@{\/}lr@{\/}lr@{\/}lr@{\/}lr@{\/}lr@{\/}l}
    $L_0$ && $T_0$ && &A,B && $\kappa$ && $\eta$ && M && $\sigma$ \\
    \hline
    36&   &  930& & 0&.083 & 0&.053& 1&.41 & 0&.1 & 0&.055\\
    \hline
    5&.9 & 71& & 0&.0625 & 0&.04 & 0&.5 & 0&.5 & 0&.042 \\
    \hline
    5&.9 & 71& & 0&.0625 & 0&.04 & 0&.5 & 0&.2 & 0&.042 \\
    \hline
    0&.95& 4&.5 & 0&.0625 & 0&.04 & 0&.2 & 0&.3 & 0&.042 \\
    \hline
    0&.15 & 0&.89 & 0&.00625 & 0&.004 & 0&.025 & 4&.0 & 0&.0042 \\
    \hline
    0&.010 & 0&.016  & 0&.00625 & 0&.004 & 0&.0065 & 2&.5 & 0 &.0042 \\
    \hline
    0&.00095 & 0&.00064 & 0&.00313 & 0&.002 & 0&.0014 & 8&.0 & 0&.0021 \\
    \hline
    0&.00030 & 0&.00019 & 0&.00125 & 0&.0008 & 0&.0005 & 10&.0 & 0&.00082 \\
  \end{tabular}
  \end{center}
\protect\label{table:pars}
\end{table}

\begin{figure}[htb!]
\begin{center}
\leavevmode
\begin{minipage}{0.48\textwidth}
        \resizebox{\textwidth}{!}{\rotatebox{-90}{\includegraphics{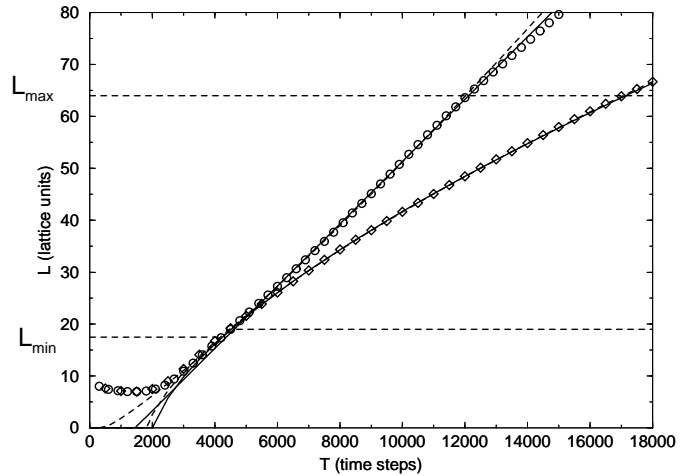}}}
	\caption{$L$ vs. $T$ for the runs shown in
		 Table \protect\ref{table:pars} with
		 $L_0$ = 5.9 (M = 0.2) (circles) and 0.0003 (diamonds).
		 The region used for fitting is delimited by 
		 ($L_{min}<L<L_{max}=64$) and the fits
		 are projected back to show the intercepts,
		 $T_{int}$. The fits (solid) are
		 to $\alpha = 1, 2/3$; free exponent fits are also shown
		 (dashed), with best fit values $\alpha = 1.16, 0.69$.}
        \protect\label{graph:raw-data}
\end{minipage}
\end{center}
\end{figure}

In all runs, the interface width is
$\xi \simeq 5 \sqrt{\kappa/2A}\simeq 3$ in lattice units.
This was found \cite{long} to be the minimum acceptable to obtain
an accurately isotropic surface tension.
To minimise diffusive effects, data for which the diffusive contribution
to the growth rate was greater than 2\% was discarded \cite{footdifcon};
this corresponded to a minimum value of $L$ of $15 < L_{min} < 24$, depending
on the run parameters.
The large size of our runs allowed a ruthless attitude to finite size
effects: we use no data with $L > \Lambda/4$, with $\Lambda$ the
linear system
size. In our $256^3$ runs, these filters mean that the good data from any
single run lies within $20 \ls L \le 64$, a comparable range to previous
studies
\cite{laradji,lebowitz,jury}. Datasets of high and low $L_0$ are
well fit
respectively by $\alpha = 1$ and $\alpha = 2/3$ (see Fig.\ref{graph:raw-data}).

\begin{figure}[htb!]
\begin{center}
\leavevmode
\begin{minipage}{0.48\textwidth}
        \resizebox{0.96\textwidth}{!}{\rotatebox{-90}{\includegraphics{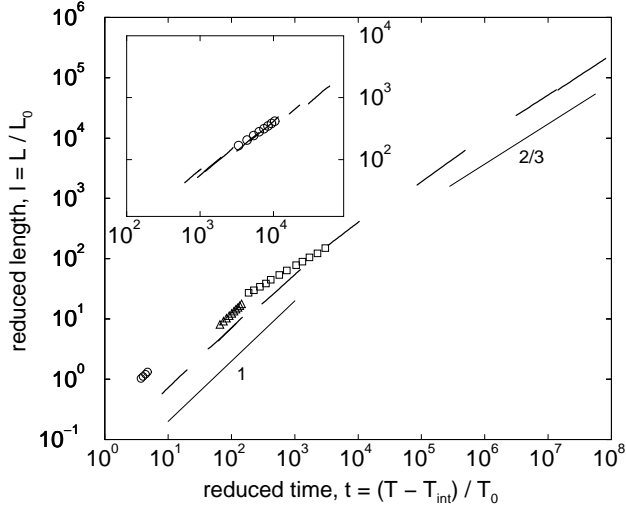}}}
        \caption{Scaling plot in reduced variables $(l,t)$ for LB data,
		 bold lines (left to right)
		 are those of Table I (top to bottom).
		 %Free exponent fits are shown dotted for the
		 %first three data sets.
		 Squares Ref.\protect\cite{foot2_3},
		 triangles Ref.\protect\cite{laradji},
		 circles Ref.\protect\cite{lebowitz}.
		 Inset: DPD data of Ref.\protect\cite{jury} (solid lines)
		 with one of our
		 data sets ($L_0 = 0.15$, circles) repeated for comparison.}
        \protect\label{graph:cooked}
\vspace{-0.5cm}
\end{minipage}
\end{center}
\end{figure}

However, as emphasized by Jury {\it et al.} \cite{jury}, meaningful tests of scaling
are best made not by looking at single data sets but by combining those of
different parameter values. To this end, the good data from each run were fit to
$L=B(T-T_{int})^\alpha$, so as to extract an intercept $T_{int}$; we then
transformed the data to reduced physical units $l$ and $t$ defined above. The
exponent $\alpha$ was first allowed to float freely; this gave
reproducible values at large $l$ and $t$ (e.g., $\alpha = 0.69$ and $0.67$ for the
last two data sets in Table 1), but more scattered ones at small $l$ and $t$
($\alpha = 0.88$, $0.86$ and $1.16$ for the first three data sets).  In the latter
region the floating fit is relatively poorly conditioned; it also gives
large relative errors in $T_{int}$ (see Fig.1). In contrast, fits to
$\alpha = 1$ for these three data sets gave much better data collapse with
consistent values of $b$ ($b=0.073$, $0.073$ and $0.072 \pm 0.01$). Thus we are
confident of $\alpha = 1$ in this region. For the remaining data sets we
estimate errors in individual exponent values at around 10\% and in
reduced time $t$ around 3\% to 10\%. 
Figure \ref{graph:cooked} shows all our data sets on a single plot
using reduced variables $l$ and $t$. Such
a plot is necessarily log-log, since our data sets span seven decades in
$t$ and five in $l$, a range which
exceeds all previous studies combined.

These LB results are fully
consistent with the existence of a single underlying scaling curve $l =
l(t)$, in
which viscous ($l= bt$) and inertial ($l=ct^{2/3}$) asymptotes are
connected by a long
crossover whose breadth 
justifies our use of a single floating exponent $\alpha$ in the
fits used above to extract $T_{int}$ for each run.
Although we cannot rule it out for still larger times $t$, we see no
evidence for a further crossover to a regime with
asymptotic exponent $\alpha \le 1/2$ as demanded by Grant and Elder
\cite{grant}.

Before considering our results in more detail,
we discuss their relation to others previously published. We
restrict attention to those 3D data sets for which reliable estimates of
$L_0$ and $T_0$ exist \cite{footlin}. Datasets of Laradji {\it et al.}
\cite{laradji} and of Bastea and Lebowitz
\cite{lebowitz} are shown on Fig.\ref{graph:cooked} (fitted to $\alpha=1$
\cite{jury}).
These lie in an $l,t$ range
($1 \ls l \ls 20$) in which our own data shows viscous (linear) scaling
$[$Eq. (\ref{viscous})$]$; both data sets were claimed to confirm the linear
law by their authors, but with differing values of $b = 0.13$, 0.3.
Our own $b$ values are lower than either (see above and Fig.\ref{graph:cooked}).
As noted above,
we took special care to ensure that the diffusive contribution to
coarsening was small; we have found
that, for matching $L_0,T_0$ values, LB data sets 
similar to
those of Refs.\cite{laradji,lebowitz} can be generated using too large a
mobility $M$.  We hypothesize
therefore that both data sets have strong residual diffusion,
leading to an overestimate of $b$.
Likewise the data of Appert {\it et al.} \cite{foot2_3}, which lies in the
crossover regime of our scaling plot, asymptotes to our data from above;
this suggests that their fitted exponent $\alpha \simeq 2/3$ is too low because
of diffusion.

A  different explanation, based on a possible nonuniversality of the
physics of
topological reconnection of domains, was suggested by Jury {\it et al.}
\cite{jury}, whose dissipative particle dynamics (DPD) results also appear in
Fig.\ref{graph:cooked} (inset) \cite{wagner}. These
authors found that each data set was well fit by a linear scaling,
Eq. (\ref{viscous}),
but with a systematic increase of the $b$ coefficient upon moving from
upper right to
lower left in the scaling plot\cite{jury}. Their
alternative suggestion was that their own data, and that of
Refs.\cite{laradji,lebowitz}, were part of an extremely
broad crossover region,
$1\ls t \ls 10^4$ in reduced time.
Our LB data support the idea of a broad crossover, but
instead places it at $10^2 \ls t \ls 10^6$. Note that, unlike those
of Refs.\cite{laradji,lebowitz}, all the data sets of Jury {\it et al.} do lie
very close to
our own (Fig.\ref{graph:cooked} inset). Since the two simulation methods are
entirely different, this lends support to the idea of a universal scaling,
although the fact that each DPD run is best fit by a locally linear growth law
does not \cite{jury}. The latter could be partly due to
finite
size effects; to obtain enough data,
Jury {\it et al.} included results up to
$L = \Lambda/2$, whereas we reject all data with $L > \Lambda/4$.

The arguments of Ref. \cite{jury} involve
the intrusion of a second length scale, alongside $L_0$, which in the
LB context is the interfacial width $\xi$ (or more generally, a
molecular
scale). The ratio $h = \xi/L_0$ for real
fluids is in the range $0.05$ (water) to $10^{-7}$ (glycerol).
In simulations, $\xi$ cannot be smaller than the lattice spacing,
and the inertial region is achieved by setting $L_0 \ll 1$, so $h \gg 1$.
In this sense our interface is ``unnaturally thick'': simulation runs that
enter
the inertial regime do so directly from a diffusive one, without an intervening
viscous regime.  However, this
should not matter if  $l(t)$ follows a universal curve, as our
results (in contrast to Ref.\cite{jury}), in fact suggest.
But the microscopic length still plays an interesting role,
as follows.  As a fluid neck stretches thinner and thinner before
breaking, it shrinks laterally to the scale $\xi$; diffusion then takes over to
finish the job of reconnection.  So, although our work involves length
scales where the direct contribution of diffusion to domain growth is
negligible, we
must ensure that it is handled correctly at smaller scales.
This factor limits the accessible range of $l$ and $t$, not only at the lower
\cite{jury} but also at the upper end \cite{long}.

\begin{figure}[htb!]
\begin{center}
\leavevmode
\begin{minipage}{0.48\textwidth}
        \resizebox{\textwidth}{!}{\rotatebox{-90}{\includegraphics{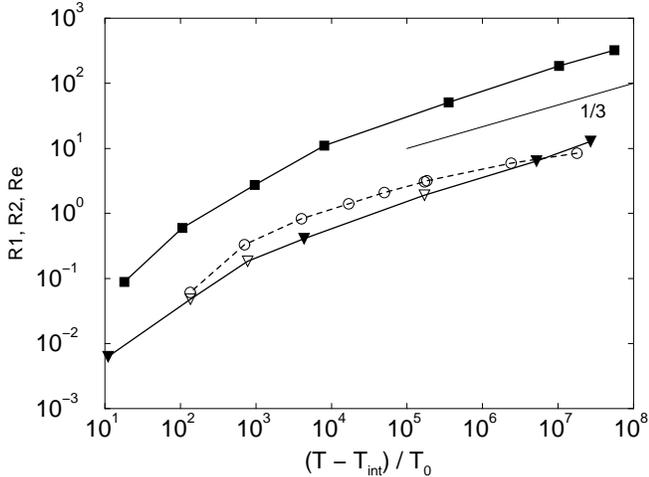}}}
        \caption{
		 Reynolds numbers Re $=l\dot{l}$ (filled squares)
		 for $L=50$ for (left to right)
		 runs in Table \protect\ref{table:pars} (top to bottom).
		 Ratios $R_1$ (circles), $R_2$ (triangles),
		 the rms inertial to viscous ratios (see text)
		 at $L=30$ for runs with (left to right)
                 $L_0$ = 36, 2.9, 0.59, 0.15, 0.054, 0.024, 0.01, 0.01, 0.0016,
		 0.00095, 0.00039, 0.0003 (system sizes $96^3$ (open) 
		 and $128^3$ (filled)).
		 Errors are of the order of the symbol size.}
        \protect\label{graph:ratios}
\vspace{-0.5cm}
\end{minipage}
\end{center}
\end{figure}

The breadth of the viscous-inertial crossover is
somewhat less extreme when expressed in terms of Re (see above);
%(Eq.\ref{reynum}); 
our data span
$0.1 \ls$Re$\ls 350$ and the crossover region is roughly $1\ls$Re$\ls 50$.
Re values
(at $L \simeq 50$) for each run are shown in Fig.\ref{graph:ratios} against
reduced
time $t$. Data are consistent with Re $\sim t^{1/3}$ as predicted from
Eq. (\ref{inertial}).
Note that, in simulating high Re flows, one should strive to ensure that the
dissipation scale \cite{turbulence}
(defined as $\lambda_d = (\eta^3/\epsilon\rho^3)^{1/4}$, with
$\epsilon$ the energy dissipation per unit volume) always remains larger
than the
lattice spacing. This ensures that any turbulent cascade (whose shortest
scale is
$\lambda_d$) remains fully resolved by the grid. Equating dissipation with
the loss
of interfacial energy, one has
$\epsilon
\simeq d(\sigma/L)/dT$ and so, in reduced units, $\lambda_d \simeq (l^2/\dot
l)^{1/4}$. Comparable $\epsilon$ values are found directly from our simulated
velocity data; and $\lambda_d$ remains larger than the grid size for all
our runs
\cite{footren}.

A decisive check that we really are simulating a regime
where inertial forces dominate over viscous ones, is based directly on the
velocity
fields found in our simulations
\cite{long}. From these we calculated rms values of the
individual terms in the Navier-Stokes equation ($\rho =1$),
$\left({\partial {\mathbf v}}/{\partial t} +
{\mathbf v}\cdot\nabla{\mathbf v}\right)
= \eta\nabla^2{\mathbf v} - \nabla\cdot{\mathbf{P}}.
$
Here ${\mathbf{P}}$, the pressure tensor, contains
the driving terms arising from interfacial tension.
Ratios $R_1 = \langle \partial
{\mathbf v}/\partial t\rangle_{rms}/\langle \eta \nabla^2
{\mathbf v}\rangle_{rms}$
and $R_2 =\langle
{\mathbf v}\cdot\nabla{\mathbf v}\rangle_{rms}/\langle \eta \nabla^2
{\mathbf v}\rangle_{rms}$,
were then computed; these can be seen in Fig.\ref{graph:ratios}.
The ratio $R_2$ is closely related to the Reynolds number Re:
it differs in representing length and velocity measures based on the rms fluid
flow rather
than on the interface dynamics and, because the length scales associated
with the velocity gradients are smaller than the domain size,
is significantly smaller than Re. The dominance (by a factor
ten)
of inertial over viscous forces is, at late times, nonetheless clear
(Fig.\ref{graph:ratios}).

We finally ask whether, at the largest Re values we can reach, there is in
fact significant turbulence in the fluid flow.  One quantitative signature of
turbulence is the skewness
$S$ of the longitudinal velocity derivatives; this is close to zero in
laminar flow but approaches $S=-0.5$ in fully developed turbulence
\cite{turbulence}.
We do detect increasingly negative $S$ as Re is increased but reach only
$S\simeq -0.3$ for Re $\simeq 350$ \cite{long}.   This suggests that at our
highest Re's, turbulence is at most
partially developed -- a view confirmed by visual inspection of velocity maps
\cite{long}. Grant and Elder's suggestion of an eventual transition to
turbulent
remixing thus remains open.

In conclusion, we have presented LB simulation data for 3D spinodal
decomposition which spans an unprecedented range of reduced time and length
scales.
At $t\ls 10^2$ (Re $\ls 1$) we observe linear scaling, as announced in the
previous
literature
\cite{laradji,lebowitz,jury,footlin}. This is followed by a long crossover
($10^2 \ls t \ls 10^6$, or $1 \ls$Re$\ls 50$)
connecting to a regime in which inertial
forces clearly dominate over viscous ones (see Fig.\ref{graph:ratios}); our
work is
the first to unambiguously probe this regime in 3D\cite{foot2_3}.
In the region so far accessible
($10^6 \ls t \ls 10^8$, or $50 \ls$Re$\ls 350$)
Furukawa's prediction of $t^{2/3}$ scaling is obeyed, to within
simulation
error. An open issue is whether this regime marks the final asymptote or
whether a
further crossover occurs to a turbulent remixing regime (saturating Re) as
proposed by Grant and Elder \cite{grant}. If it does, we have shown that
any limiting
value of Re must significantly exceed their estimate of $10-100$.

We thank Craig Johnston, Simon Jury, David McComb,
Patrick Warren and Julia Yeomans for valuable
discussions. Work funded in part under the EPSRC E7 Grand Challenge.


\begin{thebibliography}{99}
\bibitem{footdifcon} The diffusive contribution for each run was found by
running a
simulation with the same $M$, but extremely high viscosity; the data were
found to fit
well to diffusive scaling ($L = d (T-T_o)^{1/3}$) and the relevant
diffusive growth
rate $\dot L_D$ found as $\dot L_D =  d (T-T_o)^{-2/3}/3 =
d^3/3L^2$; data from the main runs with $\dot L_D/\dot L > 0.02$ was then
rejected.
(Repeating the analysis with a threshold of 1\%
did not alter the fit parameters beyond estimated
errors.)
\end{thebibliography}
\end{document}